\documentclass{moriond}

\usepackage{graphicx}
\usepackage{subcaption}

\usepackage[usenames,dvipsnames,svgnames,table]{xcolor}

\def\be{\begin{equation}}
\def\ee{\end{equation}}
\def\bea{\begin{eqnarray}}
\def\eea{\end{eqnarray}}

\newcommand{\Photo}{}
\usepackage{booktabs}
\usepackage{amsmath}
\usepackage{graphicx,color}
\usepackage{multirow}
\begin{document}
\vspace*{4cm}
\title{On the coverage of electroweak-inos within the pMSSM with SModelS - a comparison with the ATLAS pMSSM study}

\author{Leo Constantin\,\footnote{On behalf of the SModelS collaboration.}}

\address{Laboratoire de Physique Subatomique et de Cosmologie, Université Grenoble-Alpes,\\ 53 Avenue des Martyrs, F-38026 Grenoble, France}

\maketitle

\abstracts{The ATLAS collaboration has recently performed a vast scan of the phenomenological Minimal Supersymmetric Standard Model (pMSSM) with a
focus on the electroweak-ino sector, and analysed how their Run 2 searches
for electroweak production of supersymmetric (SUSY) particles constrain this
dataset. All the SLHA files from the scan as well as the constraints from the
eight individual searches considered by ATLAS were made publicly available.
We use this material to study how well the ATLAS constraints can be re-
produced with SModelS~v3.0. Moreover, we explore how the picture changes
when also including CMS results, and what can be gained by the statistical
combination of analyses. Finally, we discuss the part of parameter space with
light electroweak-inos that remains valid despite the stringent LHC limits. Our
results underscore the need of a broad, multifaceted approach for maximising
sensitivity and closing loopholes in the extensive SUSY parameter space.}

\section{Introduction}
Searches for new physics at the LHC are usually performed in specific final states and interpreted in the context of effective Simplified Models (SMS) assuming a specific set of particles and decay modes. In the absence of evidence of new physics, re-interpreting these results has become important for constraining broader and more complex Beyond the Standard Model (BSM) scenarios. The SModelS software~\cite{Altakach:2024jwk} offers a powerful framework for this endeavour, by matching the signatures of BSM models to the collection of SMS results published by the ATLAS and CMS experiments. SModelS comes with a large database (117 SMS topologies, 125 analyses in v3.0.0) allowing to investigate many different experimental signatures. The direct re-use of SMS results makes SModelS very fast and thus well suited for large scans of parameter space, such as in the context of supersymmetric models. The recent ATLAS pMSSM study~\cite{ATLAS:2024qmx} 
offers a perfect ground to test SModelS and its coverage of the electroweak-ino (EWKino) sector~\cite{Constantin:2025bqp}.

\section{Comparing SModelS with reinterpretation within the experiments }

\subsection{The ATLAS pMSSM Study}
In ~\cite{ATLAS:2024qmx}, ATLAS assessed the limits on the 19 parameters pMSSM from 8 of their Run-2 searches at full luminosity. They generated a scan with a specific focus on EWKino production, allowing for bino-like, wino-like and higgsino-like Lightest SUSY Particles (LSPs). These searches are either implemented or do have an earlier equivalent (for 36~fb$^{-1}$) in SModelS as presented in Table~\ref{tab:ana_smodels}. ATLAS provides their complete scan together with detailed information on the constraints for each point, which gives us the opportunity to test SModelS against these ``official" results.

\begin{table}[!t]
  \centering
  \caption[]{Summary of the analyses considered in \cite{ATLAS:2024qmx} and their availability in SModelS~v3.0.0.} 
  \label{tab:ana_smodels}\vspace*{-2mm}
  {\small
  \begin{tabular}{lclcc}
    \toprule
    Type   & \qquad &  ATLAS ID & \qquad & In SModelS \\
    \toprule 
    0 lepton (fully hadronic) & & \href{https://atlas.web.cern.ch/Atlas/GROUPS/PHYSICS/PAPERS/SUSY-2018-41/}{SUSY-2018-41} & & YES \\ 
     1 lepton + 2 $b$-jets & & \href{https://atlas.web.cern.ch/Atlas/GROUPS/PHYSICS/PAPERS/SUSY-2019-08/}{SUSY-2019-08} & & YES \\
    2 leptons + 0 jets & & \href{https://atlas.web.cern.ch/Atlas/GROUPS/PHYSICS/PAPERS/SUSY-2018-32/}{SUSY-2018-32} & & YES \\
     2 leptons + jets & & \href{https://atlas.web.cern.ch/Atlas/GROUPS/PHYSICS/PAPERS/SUSY-2018-05/}{SUSY-2018-05}  & & YES \\
    3 leptons on-shell  & & \multirow{2}*{\href{https://atlas.web.cern.ch/Atlas/GROUPS/PHYSICS/PAPERS/SUSY-2019-09/}{SUSY-2019-09}} & & \multirow{2}*{YES} \\
    3 leptons off-shell & &  & &  \\ 
     4 leptons & & \href{https://atlas.web.cern.ch/Atlas/GROUPS/PHYSICS/PAPERS/SUSY-2018-02/}{SUSY-2018-02} & & \href{https://atlas.web.cern.ch/Atlas/GROUPS/PHYSICS/PAPERS/SUSY-2017-03/}{SUSY-2017-03} \\
    Compressed (soft leptons) & & \href{https://atlas.web.cern.ch/Atlas/GROUPS/PHYSICS/PAPERS/SUSY-2018-16/}{SUSY-2018-16} & & YES \\ 
     Disappearing Track & & \href{https://atlas.web.cern.ch/Atlas/GROUPS/PHYSICS/PAPERS/SUSY-2018-19/}{SUSY-2018-19} & & \href{https://atlas.web.cern.ch/Atlas/GROUPS/PHYSICS/PAPERS/SUSY-2016-06/}{SUSY-2016-06}\\ 
\bottomrule
\end{tabular}}
\end{table}

\subsection{Our setup}
For our study~\cite{Constantin:2025bqp}, we re-used the ``EWKino" scan data  provided by ATLAS (12\,280 points). We removed points declared ``filtered" by ATLAS, as no sensitivity is expected for them, as well as points excluded by the upper limit on BR(h$\rightarrow$invisible) and/or the lower limit on the CP-odd Higgs boson mass, as SModelS cannot account for those. 
Additionally, we identified and removed erroneous scan points for which the mass gap $m_{\tilde\chi^{\pm}_{1} }-m_{\tilde\chi^{0}_{1}}$ is below the 2-loop mass splitting for pure winos,\footnote{This is due to a bug in the 4.0.5beta version of SPheno used in the ATLAS study.} which is the most extreme scenario for the chargino lifetime.
We end up with a total of 8\,953 points, for which we computed the production cross sections at $\sqrt{s}=8$ and 13~TeV at next-to-leading order (NLO) with Resummino, and for which we compare the SModelS~v3.0.0 constraints to the ones from the ATLAS pMSSM study. 

\subsection{Comparison}

\begin{figure}[t]    \centering
    \includegraphics[width=0.54\linewidth]{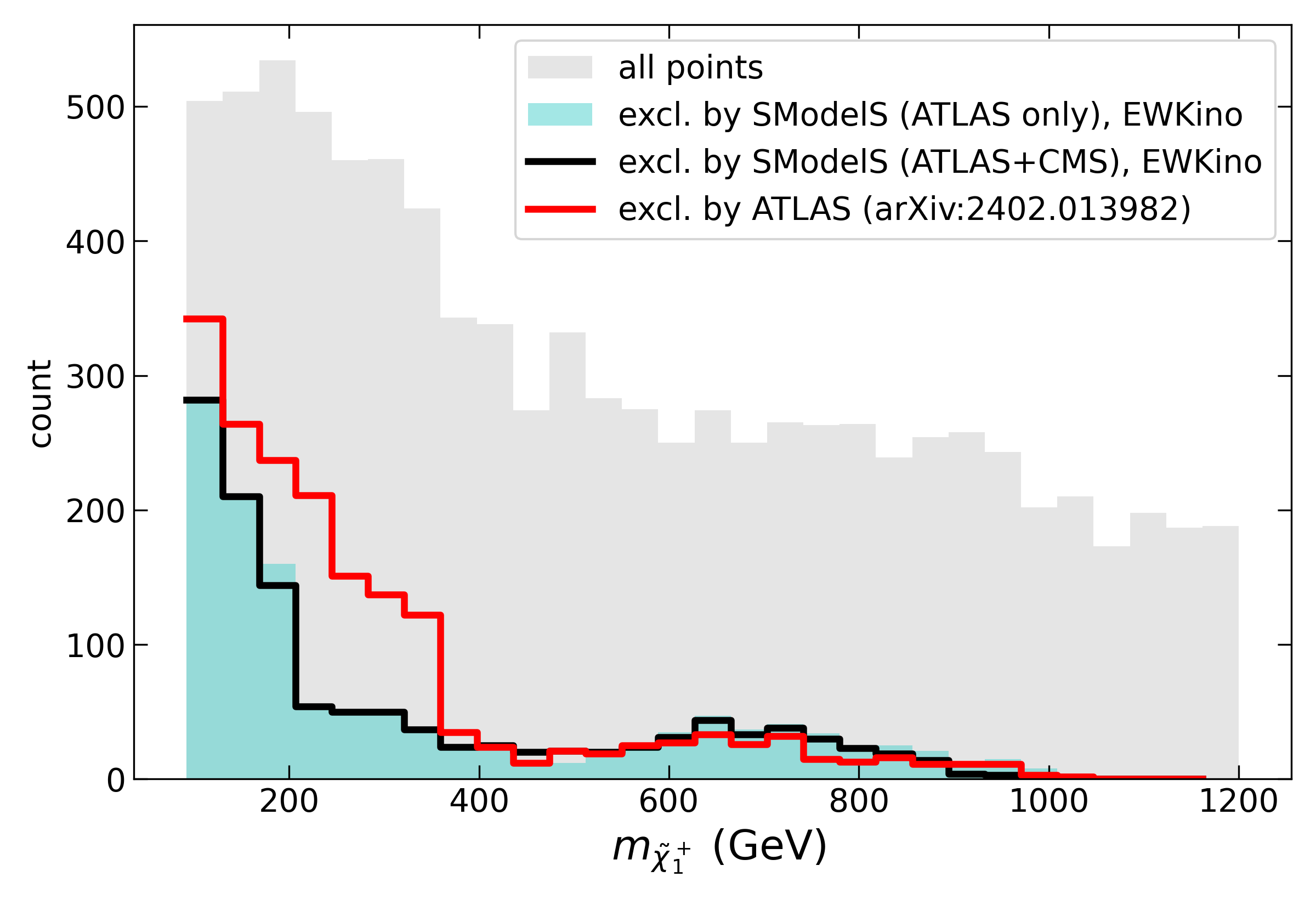}\vspace*{-3mm}
    \caption[]{Exclusion by the most sensitive analysis in SModelS, considering only ATLAS EWKino results in the database (cyan) or considering all ATLAS+CMS EWKino results (black). The open red histogram shows the corresponding official exclusion \cite{ATLAS:2024qmx} from ATLAS. The full dataset is represented in grey.}
    \label{fig:histoex}
\end{figure}

Figure~\ref{fig:histoex} shows the number of points excluded by ATLAS and by SModelS as a function of the lightest chargino mass $\tilde{\chi}_{1}^{\pm}$. For each scan point, only the constraint from the most sensitive EWKino analysis is considered. 
We observe good agreement between ATLAS and SModelS, especially for $m_{\tilde\chi^{\pm}_{1} }$ above 360 GeV, with a slight over-exclusion from the fully hadronic search, SUSY-2018-41, in SModelS. 
For masses below 360 GeV, SModelS excludes less points than ATLAS. This difference arises mainly from the disappearing track signature, for which the SModelS~v3.0.0 database contains only the 36~fb$^{-1}$ results compared to the 136~fb$^{-1}$ ones used in the ATLAS study. 
This concerns about 60\% of the under-excluded points.
Moreover,  
in scenarios with higgsino-like or wino-like LSPs, some of the signal can be missed in SModelS due to a lack of $Wh+$MET (missing transverse energy) and $ZZ+$MET efficiency-map results 
from the 3~leptons analysis SUSY-2019-09. In a similar trend, in the scheme of SModelS, some multi-step decays do not have any SMS results to be matched onto, which weakens the constraining power of the 2 leptons + jets search SUSY-2018-05.

Overall, we are able to track and identify the gaps in the coverage  
as coming from the diversity of topologies and the material available for the implementation of searches in the SModelS database.
CMS EWKino results, being generally in agreement with the ATLAS ones, do not change the picture. 

\section{Full database and analysis combination}

\begin{figure}[t]
\centering
\begin{minipage}[t]{.48\textwidth}
  \centering
  \includegraphics[height=5.2cm]{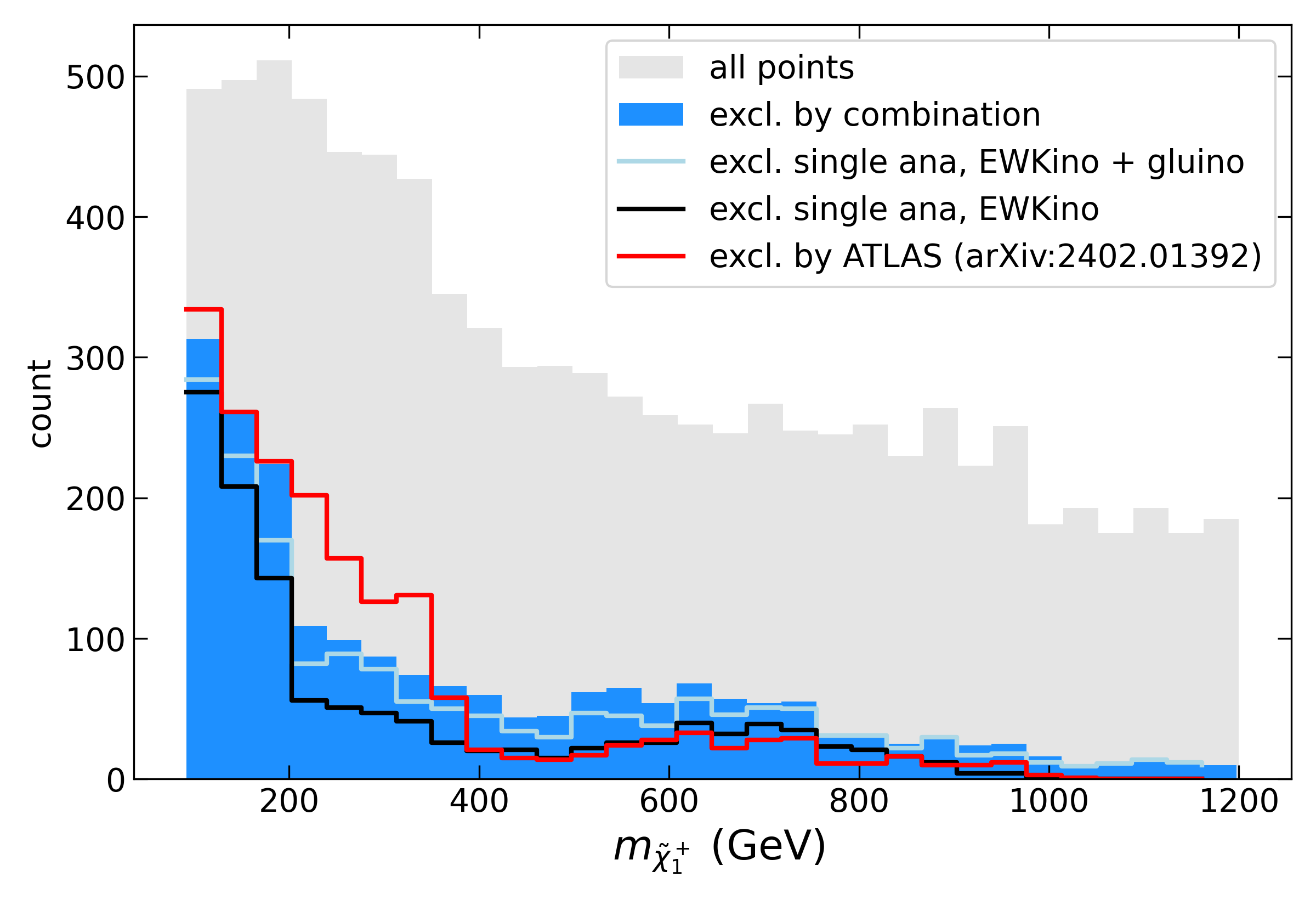}
  \captionof{figure}[]{Number of points excluded by SModelS in different setups discussed in the text. For comparison, the red histogram shows the exclusion from ATLAS~\cite{ATLAS:2024qmx}. The full dataset is represented in grey.}
  \label{fig:histocomb}
\end{minipage}%
\hfill
\begin{minipage}[t]{.48\textwidth}
  \centering
  \includegraphics[height=5.7cm]{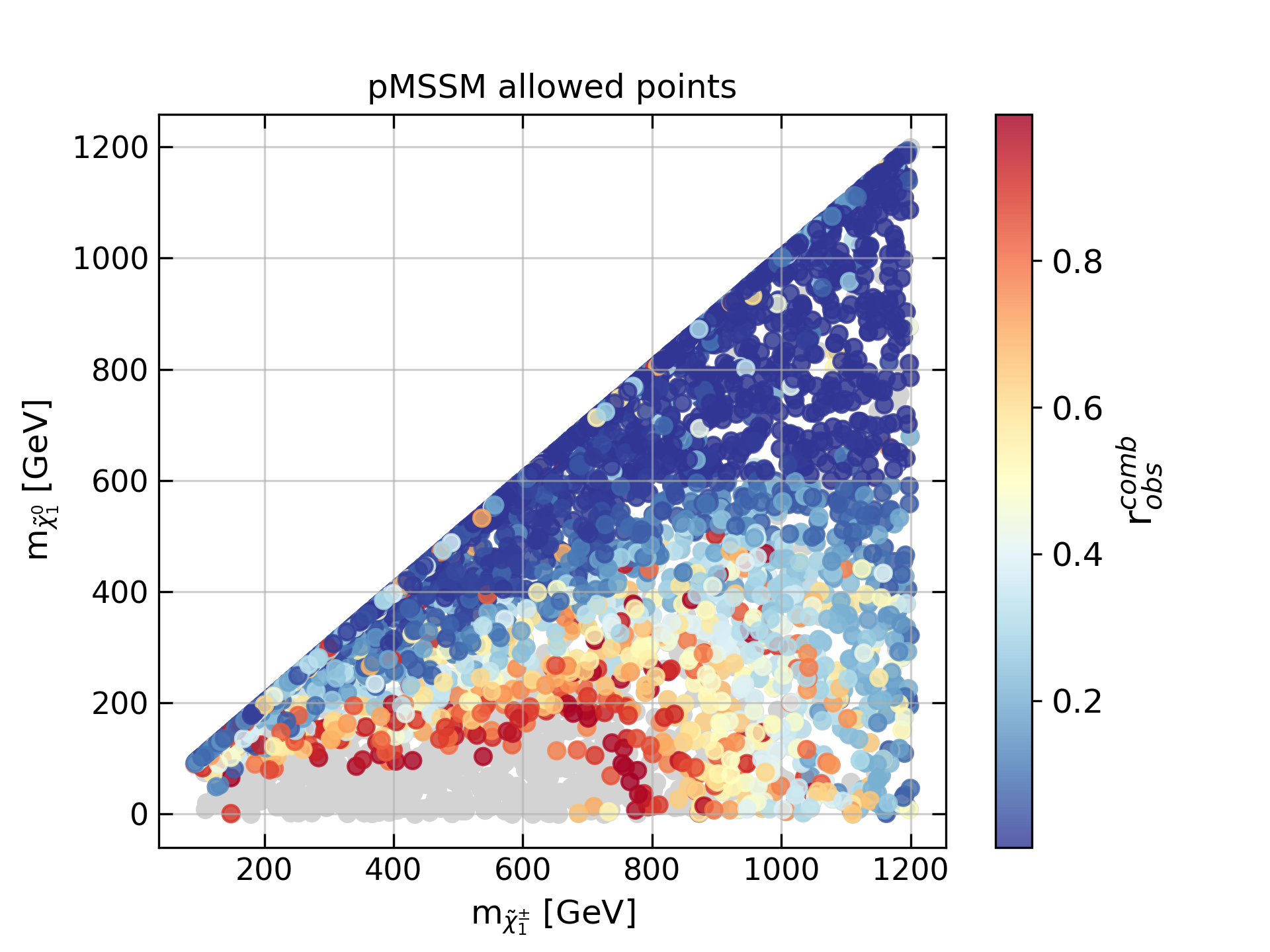}
  \captionof{figure}{Scatter plot of pMSSM scan points that escape exclusion (in colour). Grey points are excluded by the combination of analyses. The ATLAS disappearing track results for  140~fb$^{-1}$ are also accounted for.}
    \label{fig:ewinoAllowed}
\end{minipage}
\end{figure}

Let us now go a step further and use the full potential that SModelS has to offer to constrain this dataset. 
To this end, first note that the EWKino scan generated by ATLAS includes also points with light gluinos. 
We thus added the gluino pair production cross sections at NNLL, computed with NLLfast, and considered also gluino constraints in the SModelS database. The result is shown as light blue line in Fig.~\ref{fig:histocomb}. There is a clear improvement with respect to considering only EWKino results (black line) over the full range of chargino masses. 

Additionally, it is also possible to combine uncorrelated analyses in SModelS. 
 This allows one to incorporate results from multiple channels and/or experiments in a single likelihood, thus increasing sensitivity while reducing statistical fluctuations. 
Following the method of \cite{Altakach:2023tsd}, we consider up to 37 analyses (24 from ATLAS, 13 from CMS) for combination in our study, retaining for each scan point the combination that maximizes sensitivity. The outcome is shown as blue histogram in Fig.~\ref{fig:histocomb}, again depicting a significant improvement over the previous results discussed here. 

Table~\ref{tab:excl} summarizes the increase in exclusion at each step of our study. If the coverage of ATLAS and CMS is roughly the same, adding gluino constraints markedly increases the coverage of the parameter space. 
The optimal combination of uncorrelated analyses further strengthens the constraints by leveraging more of the available data. This is especially important for scenarios that feature a variety of different signatures, and are thus constrained by several analyses which each consider distinct final states. 

\begin{table}[h]
\centering
\caption{Comparison of the number of excluded points for bino-like LSP and non-bino-like LSP. Note that for single-analysis results (cases 1--4), the most sensitive, not the most constraining, analysis is considered.}
\label{tab:excl}\vspace*{-3mm}
{\small
\begin{tabular}{l|c|c}
 & bino-like LSP & non-bino LSP \\
 \hline
Total number of points after filtering & 3034 & 5919 \\
\hline
Number of points excluded by: & & \\
\quad 1. ATLAS in arXiv:2402.01392  & 529 (17\%) & 1271 (21\%)\\
\quad 2. ATLAS EWKino results in SModelS &575 (19\%) & 687 (12\%) \\
\quad 3. all EWKino results in SModelS & 539 (18\%) & 662 (11\%)\\ 
\quad 4. full SModelS database & 666 (22\%) & 1030 (17\%)\\
\quad 5. full DB with analysis combination & 847 (28\%) & 1184 (20\%)\\
\bottomrule
\end{tabular}
}
\end{table}

From these results it is clear that, contrary to some beliefs, light SUSY states are by no means ruled out~\cite{Constantin:2025mex}. An attempt at characterizing the allowed and  excluded parameter regions is made in Fig.~\ref{fig:ewinoAllowed}. Shown is a scatter of the pMSSM scan points in the plane of lightest neutralino $\tilde\chi^0_1$ versus chargino $\tilde\chi^\pm_1$ mass. 
Points in grey are excluded by the global likelihood from the analysis combination; as one can see, this concerns only a small region with charginos below about 700~GeV and bino-like LSPs below about 100~GeV. Points in colour remain allowed.\footnote{Points allowed by the combination but excluded in \cite{ATLAS:2024qmx} by the disappearing tracks analysis with 136~fb$^{-1}$, for which only the 36~fb$^{-1}$ results are available in SModelS, have been removed.} 

The colour code shows the combined $r$-value from SModelS ($r=\frac{\sigma^{\text{BSM}}}{\sigma_{95}}$ being the ratio of the predicted signal cross section over its 95\% CL upper limit), indicating how close or far a point is from exclusion. From na\"ive rescaling one can expect future Run~3 results to probe red-orange-yellow points. Even so, light EWKinos below the TeV scale, especially light higgsinos as motivated by natural SUSY, remain a viable possibility; more details are discussed in~\cite{Constantin:2025mex}.

\section{Conclusions}
As LHC operation continues and the volume of experimental data expands, the repurposing of the LHC results to explore novel BSM scenarios
---whether previously untested or not yet theorized---represents a logical and compelling approach for the community. 
In this context, SModelS emerges as a convenient and efficient public tool
, enabling users to quickly assess the viability of new physics scenarios. 

We have shown in the context of the pMSSM, that SModelS is well suited to tackle large scans and that its results are in good agreement with experiment-internal reinterpretations. We also  demonstrated that analysis combination has a significant effect on the constraining power, with a 20\% (35\%) increase in (expected) exclusion. SModelS relies on the public material provided by the experiments. We thus welcome ATLAS and CMS efforts to provide efficiency maps for the relevant simplified models.


\section*{References}
\bibliography{constantin}

\end{document}